\documentclass[letterpaper]{article} 
\usepackage{aaai24}   
\usepackage{times}  
\usepackage{helvet}  
\usepackage{courier}  
\usepackage[hyphens]{url}  
\usepackage{graphicx} 
\usepackage{natbib}  
\usepackage{caption} 
\usepackage{algorithm}
\usepackage{algorithmic}
\usepackage{multirow}

\usepackage{xcolor}

\usepackage{newfloat}
\usepackage{listings}
\DeclareCaptionStyle{ruled}{labelfont=normalfont,labelsep=colon,strut=off} 
\floatstyle{ruled}
\newfloat{listing}{tb}{lst}{}
\floatname{listing}{Listing}
\title{Cross-Platform Short-Video Diplomacy: Topic and Sentiment Analysis of China–US Relations on Douyin and TikTok}

\title{Cross-Platform Short-Video Diplomacy: Topic and Sentiment Analysis of China–US Relations on Douyin and TikTok}

\author {
    Zheng Wei\textsuperscript{\rm 1},
    Mingchen Li\textsuperscript{\rm 2},
    Junxiang Liao\textsuperscript{\rm 3},
    Zeyu Yang\textsuperscript{\rm 2},
    Xiaoyu Yang\textsuperscript{\rm 2},
    Yixuan Xie\textsuperscript{\rm 2},
    Pan Hui\textsuperscript{\rm 2},
    Huamin Qu\textsuperscript{\rm 1}
}
\affiliations {
    \textsuperscript{\rm 1}The Hong Kong University of Science and Technology\\
    \textsuperscript{\rm 2}The Hong Kong University of Science and Technology (Guangzhou)\\
    \textsuperscript{\rm 3}The Hong Kong Polytechnic University\\
    zwei302@connect.ust.hk, mli736@connect.hkust-gz.edu.cn, junxiang.liao@connect.polyu.hk, zyang979@connect.hkust-gz.edu.cn, 
    xyang058@connect.hkust-gz.edu.cn, yixuanxie@hkust-gz.edu.cn, panhui@hkust-gz.edu.cn,
    huamin@cse.ust.hk
}

\usepackage{bibentry}

\begin{document}

\maketitle

\begin{abstract}
We examine discussions surrounding China-U.S. relations on the Chinese and American social media platforms \textit{Douyin} and \textit{TikTok}. Both platforms, owned by \textit{ByteDance}, operate under different regulatory and cultural environments, providing a unique perspective for analyzing China-U.S. public discourse. This study analyzed 4,040 videos and 338,209 user comments to assess the public discussions and sentiments on social media regarding China-U.S. relations. Through topic clustering and sentiment analysis, we identified key themes, including economic strength, technological and industrial interdependence, cultural cognition and value pursuits, and responses to global challenges. There are significant emotional differences between China and the US on various themes. Since April 2022, the Chinese government has implemented a new regulation requiring all social media accounts to disclose their provincial-level geolocation information. Utilizing this publicly available data, along with factors such as GDP per capita, minority index, and internet penetration rate, we investigate the changes in sentiment towards the U.S. in mainland China. This study links socioeconomic indicators with online discussions, deeply analyzing how regional and economic factors influence Chinese comments on their views of the US, providing important insights for China-U.S. relationship research and policy making.
\end{abstract}



\section{Introduction}
The dynamic relationship between China and the United States has been a focal point in global politics, economics, and international relations. As these two superpowers navigate the complexities of cooperation and competition, social media platforms such as the American \textit{TikTok} and the Chinese \textit{Douyin} have introduced new dimensions to their interactions. Although both platforms are owned by \textit{ByteDance}, they operate under different regulatory frameworks and cultural contexts, which not only shape the user experience but also serve as channels for public discourse that reflect and influence mutual perceptions between the two countries. Millions of users engage in discussions and share content on these platforms, offering real-time insights into societal attitudes. These interactions provide a rich data source for analyzing how digital dialogues construct and mediate national identities, stereotypes, and geopolitical narratives. By examining user comments on short videos related to each other’s content on \textit{TikTok} and \textit{Douyin}, this study aims to uncover the core themes and emotional trends in public discourse surrounding China-U.S. relations.

Furthermore, we conduct an in-depth analysis of the emotional differences towards the United States across various regions in China. By utilizing \textit{Douyin}'s IP geolocation mandatory disclosure policy implemented in April 2022, we associate user comments with their geographic locations to analyze emotional differences among different provinces. This not only enhances the comprehensiveness of public sentiment analysis but also overcomes the sample selection bias caused by users' voluntary location disclosure. By integrating geographical information, we further investigate how regional socioeconomic factors (such as per capita GDP, minority index, and internet penetration rate) influence public discussions on these platforms. Additionally, by using large language models (LLM) for sentiment analysis, we quantify the emotional tones in the comments, categorizing them into positive (including 1.rational support 2. extreme enthusiasm) and negative (including 1.hate 2.criticism or complaints), as well as neutral sentiments. This approach provides a more nuanced understanding of the emotional trends present in the discussions.

We analyze comments posted by \textit{Douyin} and \textit{TikTok} users from China and the United States on short videos involving China-U.S. relations, focusing on: (1) the topics discussed by users from both countries regarding each other's views in the comments; (2) the level of negative sentiment in the United States towards China; (3) the level of negative sentiment in comments from different provinces in China towards the United States; (4) the socioeconomic factors that can explain the emotional differences in discussions across different provinces in China. We analyzed data from 338,209 comments on 4,040 short videos displaying content related to China and the United States to answer the following questions:
\begin{itemize}

\item{\textbf{RQ1: What are the dominant themes that emerge in user comments on U.S. \textit{TikTok} and Chinese \textit{Douyin} when discussing mutual perceptions between the two countries?}} 

\item{\textbf{RQ2: Do sentiment patterns vary across different themes related to perceptions of China and the U.S., and if so, what are the key differences?}}

\item{\textbf{RQ3: Are there regional variations in the topics and sentiment attitudes expressed in comments on \textit{Douyin} in mainland China when discussing the U.S., and if so, what are the characteristics of these variations?}}

\item{\textbf{RQ4: What factors contribute to the differences in negative sentiment levels towards the U.S. across various topics and provinces in China, and can these factors be used to explain the observed variations?}} 

\end{itemize}

We found that sentiments in China towards the U.S. are relatively complex, with 11.19\% of the comments being positive and 45.55\% negative. In contrast, 42.09\% of comments from the U.S. towards China are positive, while 11.36\% are negative. Through topic clustering, we identified key themes such as economic strength, technological competition and cooperation, happy life, and addressing global challenges. Discussions in China comments tend to emphasize power dynamics and the complexity of interdependence, while U.S. comments discussions focus more on shaping values and perceptions. Based on provincial regression data, we also found that the level of negative sentiment toward China-U.S. relations varies across different regions in China. By examining socioeconomic development indicators in several provinces, including local GDP, minority population indices, and internet penetration rates, we can explain some of these regional differences. This study deepens our understanding of discussions on China-U.S. relations on social media, revealing how these discussions are influenced by regional socioeconomic development and evolve accordingly.

\section{BACKGROUND AND RELATED WORK}

Social media platforms, such as TikTok and Douyin, have significantly influenced the dynamic development of cross-cultural perceptions and international relations, providing unique digital spaces to observe the socioeconomic, political, and cultural environments of their respective user bases \cite{chu2011electronic, baum2019media, bennett2012personalization,wei2024digital}. China-U.S. relations are a critical topic in contemporary international affairs, with interactions between the two nations profoundly affecting the global political and economic landscape \cite{chengqiu2020ideational}. A substantial body of research has highlighted the significant role of public attitudes in shaping bilateral relations, often serving as references for national foreign policy formulation \cite{qin2023did, hudson2005foreign, holsti2009public}. In this context, numerous studies have focused on the perceptions and attitudes between China and the United States, typically using opinion polls and social surveys as their primary methodologies \cite{swedlow2009value, king2004enhancing}. Specifically, research on Chinese public attitudes toward the U.S. reveals a spectrum of complex sentiments often influenced by changes in international relations \cite{deng2001hegemon}. The emotional dynamics between China and the U.S. are shaped by geopolitical factors and domestic influences in both countries \cite{deng2001hegemon}. In the United States, anxiety surrounding China's rise is a significant concern \cite{ikenberry2008rise}, stemming partly from a perceived “mystique” surrounding China and partly from worries about America's own developmental trajectory \cite{friedberg2005future}.

This section reviews research on sentiment trends between China and the United States on social media. We then briefly introduce policies related to IP geolocation disclosure on Chinese social media platforms \cite{zhou2024can,wei2024social}. Finally, we present the theoretical model underpinning this study.

\subsection{Mutual Perceptions on TikTok and Douyin}

Social media has emerged as a crucial space for shaping and articulating inter-nation perceptions \cite{ausat2023role,duncombe2019politics}. \textit{TikTok} and \textit{Douyin}, both owned by \textit{ByteDance}, share a unified algorithmic architecture and similar operational models, providing a consistent foundation for comparative analysis. This commonality in their technological and usage frameworks sets the stage for examining how these platforms diverge when situated within distinct cultural and regulatory environments, resulting in varied content ecosystems. Such a setup offers a valuable opportunity to investigate the impact of localized factors on platform usage and content generation \cite{kaye2021co,liu2022dual}. Studies have shown that \textit{TikTok} users in the United States engage with a diverse range of content that contributes to a global youth culture characterized by viral trends, creative challenges, and meme sharing \cite{boffone2022tiktok,boffone2022introduction}. In contrast, \textit{Douyin} (the Chinese version of \textit{TikTok}) offers content that is more localized and closely aligned with Chinese cultural norms, social issues, and adheres to government regulations \cite{wagner2023tiktok,chen2021positiveenergy}. \textit{Douyin} emphasizes themes that reinforce Chinese cultural identity and societal values, resulting in a user experience deeply rooted in local culture \cite{chen2023my}. Previous studies have explored how these platforms reflect and shape mutual perceptions. Some scholars have analyzed user-generated content during significant geopolitical events, finding that users often engage in discourses reinforcing national identities and stereotypes \cite{zhang2019right}. The algorithmic curation on these platforms can create echo chambers, reinforcing existing beliefs and potentially exacerbating misunderstandings between cultures \cite{raza2024algorithmic,wei2024digital,brady2023algorithm}. Additionally, the lack of direct interaction between \textit{TikTok} and \textit{Douyin} users due to platform segregation limits opportunities for cross-cultural engagement.

\subsection{Sentiment Differences Across Themes}
Sentiment analysis on social media has revealed that public attitudes toward foreign nations vary across different themes and contexts \cite{thelwall2018social,weicontextaware}. In the context of China-U.S. relations, sentiments expressed fluctuate depending on topics such as politics, economics, culture, or technology \cite{zhang2020understanding,guan2020chinese,chen2022war}. During the COVID-19 pandemic, conspiracy theories and misinformation blaming China proliferated on social media, including \textit{TikTok}, fueling xenophobia and anti-Asian incidents \cite{shahsavari2020conspiracy,ittefaq2022pandemic,roberto2020stigmatization}. Moreover, concerns over data security and privacy tied to Chinese apps have deepened skepticism and distrust among American users \cite{kokas2018platform,liu2021rise}. Debates surrounding the potential banning of \textit{TikTok} in the U.S. highlighted broader issues of digital security and international rivalry \cite{ke2023unraveling,gray2021geopolitics}. These examples underscore the critical role that thematic context plays in shaping public sentiment toward foreign countries on social media platforms, demonstrating how different topics can either bridge cultural gaps or exacerbate existing tensions.

\subsection{Regional Sentiment Distribution in China}
China’s regional diversity raises questions about whether users from different provinces prioritize distinct topics or express varying sentiments on China-U.S. relations. Prior research has shown regional patterns in areas such as public health, sports, and urban planning based on users’ disclosed locations \cite{baucom2013mirroring,milusheva2021applying,cao2018using,kuchler2022jue}. However, voluntary location sharing can introduce selection bias \cite{li2016privacy,bao2015recommendations,backstrom2010find}. To address this, the Chinese government mandated that social media platforms automatically tag comments with users’ IP-based provincial locations (or country if international), minimizing self-selection concerns \cite{wei2024social}. This policy supports a more comprehensive analysis of regional sentiment variations in discussions about China-U.S. relations.

\subsection{Theoretical Models}
We propose a four-dimensional theoretical framework that integrates power transition theory, complex interdependence theory, constructivism, and global governance theory to guide our understanding of the dimensions of topics discussed between China and the U.S. on social media.

\textbf{Power Transition Theory}
Power transition theory posits that the international system is characterized by a hierarchy of states, where global order is contingent upon the distribution of power among major players \cite{ross2008china}. In the context of China-U.S. relations, this framework elucidates the competitive dynamics inherent in their interactions on social media. As China emerges as a formidable global power, the theoretical lens reveals a struggle for dominance that transcends mere economic competition, encompassing value battles over narrative control and legitimacy \cite{pan2012knowledge,ambrosetti2012power}. The interactions on social media can be viewed as a battleground for competing visions of international order, where both nations seek to assert their influence and reshape public perceptions, ultimately impacting diplomatic engagement and conflict resolution strategies \cite{collins2019hashtag}.

\textbf{Complex Interdependence Theory}
Complex interdependence theory offers a sophisticated lens through which to comprehend the multifaceted relationships characterizing China-U.S. interactions in an increasingly globalized context \cite{zhao2010managing}. It departs from traditional state-centric paradigms by emphasizing the importance of transnational networks, non-state actors, and multiple interaction channels—including economic interdependence, cultural exchanges, and technological partnerships \cite{chengqiu2020ideational}. On social media platforms, this interdependence is evident through shared narratives and dialogues that highlight mutual vulnerabilities and interconnected interests \cite{schindler2024second}. Examining these digital interactions facilitates a critical exploration of how collective identities are constructed and negotiated in the online public sphere, uncovering both opportunities for cooperation and the potential for misperceptions that may exacerbate tensions between the two nations.

\textbf{Constructivism}
Constructivism posits that state identities and interests are socially constructed through interactions and shared understandings, rather than being inherently fixed \cite{badea2021us,friedberg2005future}. This theoretical perspective emphasizes the significance of social cognition, cultural values, and collective meanings in shaping international relations. In the context of social media, constructivism is highly relevant as platforms like \textit{TikTok} and \textit{Douyin} facilitate the exchange of narratives and cultural expressions that contribute to the mutual construction of national identities \cite{jaworsky2021politics}. These discursive practices influence how nations perceive each other, impacting public opinion and potentially informing foreign policy decisions \cite{baum2008relationships}. By examining the ways in which users engage with content and with each other online, we can better understand how cultural perceptions and value systems shape bilateral attitudes and relationships.

\textbf{Global Governance Theory}
Our final theoretical lens, global governance theory, explores the mechanisms by which state and non-state actors collaborate to manage transnational challenges that transcend national boundaries \cite{higgott2000non,green2018transnational}. This perspective is particularly salient in the context of social media, where China and the U.S. interact with and respond to complex global issues such as climate change, public health crises, cybersecurity threats, and economic interdependence. By analyzing discussions on platforms like \textit{TikTok} and \textit{Douyin}, we can discern how each nation articulates its priorities, frames global challenges, and proposes solutions within the digital public sphere. 

\section{DATA}
Our data was sourced from \textit{Douyin} and \textit{TikTok}. \textit{Douyin} is China’s leading short video platform, boasting over 700 million daily active users \cite{RN2}. \textit{TikTok}, its global counterpart, has been downloaded more than 6.7 billion times \cite{zheng2021tiktok}. Short videos on platforms like \textit{Douyin} and \textit{TikTok} are often tagged with multiple hashtags, many of which are automatically suggested by the platform to reflect relevant topics. Our data collection process was driven by keyword searches, starting with seed keywords related to topics such as ``China'' and ``United States,'' along with their associated terms, which the platform recommends when performing the search. These seed keywords were initially selected based on expert knowledge of China-U.S. studies and were supplemented with platform-suggested terms that reflect trending topics related to the seed keywords. For example, when ``China'' was used as the seed keyword, the platform might recommend hashtags such as ``Chinese Society,'' ``Chinese Culture,'' or ``Life in China.'' Therefore, our keyword search method was primarily guided by the platform's algorithmic suggestions. This approach is consistent with previous research that gathered topic-related short videos on Douyin using a similar method \cite{wei2024social}. We also tested alternative methods, such as using cosine similarity to identify words related to the seed keywords, but found the platform’s algorithmic recommendations to be more effective. We began our data crawling with an initial set of ten seed keywords and expanded the list when the number of comments collected was insufficient. The final set of keywords included 40 terms, which provided a diverse range of videos and comments. These keywords include terms such as ``Chinese Society,'' ``Chinese Culture,'' and ``American Life,'' as specified in Appendix Table \ref{A1}. 

All video content featuring at least one of these keywords was collected. Between August 31 and September 19, 2024, we developed a customized web crawler utilizing the open-source tool \textit{MediaCrawler} and conducted data collection for 20 hours each day \footnote{\url{https://github.com/NanmiCoder/MediaCrawler}}. When collecting data from China, to minimize content discrepancies caused by the platform’s recommendation algorithms based on IP geolocation, we employed IP proxy services from \textit{chinapptp}. This allowed us to simulate IP addresses from multiple regions within China. When handling comments from the United States, given that \textit{TikTok}'s user base is global, we restricted our data collection to English and Chinese comments posted under videos containing the relevant search terms. Furthermore, we filtered these comments to include only those from accounts registered in the United States, thereby ensuring that the commenters are predominantly U.S. users. 

Our keyword searches yielded 1,546 short videos and 539,443 associated comments from China, alongside 3,571 short videos and 298,920 comments from the U.S., covering the period from May 2019 to September 2024. The preprocessing of comment data involved several systematic steps. Initially, four researchers assessed the videos to verify their relevance to China-U.S. relations issues. We then removed duplicate comments and those consisting solely of punctuation, emojis, or @usernames, resulting in an initial refined dataset of 338,209 comments. This included 1,158 videos and 211,090 comments from China, and 2,882 videos and 127,117 comments from the U.S. These comments were subsequently analyzed thematically. Each comment was categorized into multiple themes based on its content (refer to the ``Topic Analysis'' section below). In addition, when analyzing regional differences within mainland China, we excluded comments without IP addresses (i.e., comments posted before the mandatory IP address disclosure policy) and comments with IP addresses not from mainland China. Ultimately, 330,670 comments from 4,024 videos were included in the analysis (China: 1,142 videos, 203,553 comments; U.S.: 2,882 videos, 127,117 comments).

\section{METHOD}

\subsection{Topic Analysis}

We identified the key themes arising from online discussions through a mixed analytical strategy that combines both inductive and deductive methods, following the principles of Grounded Theory \cite{corbin2014basics}. This approach categorizes comment texts into clusters of words, subsequently synthesizing them into broader themes and theoretical dimensions, as illustrated in the thematic analyses conducted by Boyd-Graber et al. and Baumer et al. \cite{baumer2017comparing,boyd2017applications}. Next, we applied the \textit{BERTopic} algorithm \cite{grootendorst2022bertopic} to organize a substantial dataset of unstructured user comments. The algorithm clustered 539,443 Chinese comments into 301 thematic categories and 298,920 U.S. comments into 300 thematic categories.

Subsequently, we narrowed the investigation to specific clusters directly related to China-U.S. relations. To accomplish this, three researchers manually verified the common keywords within each cluster. We removed 52 clusters from the Chinese comments that were unrelated to the themes, leaving 249 clusters corresponding to 1,158 short videos and 211,090 comments. Similarly, we deleted 69 clusters from the U.S. comments that were unrelated to the themes, resulting in 231 clusters corresponding to 2,882 short videos and 127,117 comments. This constituted the final analysis sample.

Three researchers manually reviewed each cluster to ensure consistency of keywords and accuracy of the identified themes. We conducted qualitative analyses on the 249 Chinese clusters and 231 U.S. clusters. Using an iterative, consensus-based coding procedure, three researchers independently inspected the keyword lists for every cluster, manually merged clusters with synonymous or conceptually overlapping terms, and—after two rounds of discussion to reconcile differences—consolidated them into 21 higher-order themes; for instance, clusters 37 and 102, which shared $>$ 80 \% tariff-related trade keywords, were combined as “Sino-US Trade \& Economic Frictions. Finally, we mapped the themes onto a four-dimensional theoretical framework encompassing power transition theory, complex interdependence theory, constructivism, and global governance theory. Independent theme grouping was followed by collaborative discussions to resolve discrepancies and achieve consensus. Table \ref{topic-0} presents the final classification results.

\begin{table}[ht]
\centering
\caption{The classification of 21 thematic clusters into four
theoretical dimensions}
\scalebox{0.66}{
\begin{tabular}{cccc}
\hline
\begin{tabular}[c]{@{}c@{}}Power \\ Transition \\ Theory\end{tabular}                                                                                                                                           & \begin{tabular}[c]{@{}c@{}}Complex \\ Interdependence \\ Theory\end{tabular}                                                                                                             & Constructivism                                                                                                                                    & \begin{tabular}[c]{@{}c@{}}Global \\ Governance\\  Theory\end{tabular}                                                                           \\ \hline
\begin{tabular}[c]{@{}c@{}}1.Economic Strength\\ 2.Great Power Rivalry\\ 3.Power Dismantlement\\ 4.Values Dialogue\\ 5.War and Conflict\\ 6.Political Party Rivalry\\ 7.Racial Conflict\end{tabular} & \begin{tabular}[c]{@{}c@{}}1.Technological \\ Competition \\ and Cooperation\\ 2.Financial \\ Interdependence\\ 3.Industrial Synergy\\ 4.Talent Exchange \\ and Interaction\end{tabular} & \begin{tabular}[c]{@{}c@{}}1.Happy Life\\ 2.Culture and \\ Education\\ 3.Immigration\\ 4.Public Safety\\ 5.Gender/Racial \\ Equality\end{tabular} & \begin{tabular}[c]{@{}c@{}}1.Invasive species\\ 2.Food security\\ 3.Public health\\ 4.Environmental \\ protection\\ 5.Multipolarity\end{tabular} \\ \hline
\end{tabular}
}
\label{topic-0}
\end{table}

\subsection{Sentiment Analysis}
To ensure the accuracy and consistency of sentiment annotation, we developed comprehensive annotation guidelines and published them on GitHub\footnote{\url{https://anonymous.4open.science/r/DDPP-China-US-26A2}} to ensure all participants followed standardized procedures. We randomly selected 1,000 comments each from China and the United States for manual annotation, establishing a benchmark and evaluating the performance of the \textit{Qwen-Plus} LLM\footnote{\url{https://cn.aliyun.com/product/bailian?from_alibabacloud=}} and \textit{GPT-4o mini} LLM\footnote{\url{https://platform.openai.com/docs/models/gpt-4o-mini}}. After experiments with prompt optimization and fine-tuning using manually annotated sentiment analysis of 100 comments from both China and the U.S., the LLM's sentiment analysis ability was significantly improved. We selected the best-performing configuration from all of our experiments and tested it on a total of 2,000 reviews. the results showed that this configuration achieved near-human level accuracy, and the accuracy of sentiment analysis reached 88.4\% for China and 85.9\% for the U.S., as shown in Table \ref{sentiment-test}.

\begin{table}[ht]
\centering
\caption{Sentiment distribution for China and the U.S. was evaluated with 1,000 samples each, comparing manual labels with GPT-4o mini and Qwen-Plus. Where FT stands for fine-tuning, PO stands for prompt optimization. When calling the API of the LLM, the parameters are set as top\_p = 0.1, presence\_penalty = -2.0, with other parameters using default values. During fine-tuning of the LLM, the settings are epochs = 3, batch size = 1, LR multiplier = 1.8, and seed = 1925383673. The data used for prompt optimization and fine-tuning are not included in the test data.}
\scalebox{0.7}{
\begin{tabular}{ccccccc}
\hline
Method                                                                    & Country  & Negative & Neutral & Positive & Censored & Accuracy \\ \hline
\multirow{2}{*}{\begin{tabular}[c]{@{}c@{}}Manual \\ labels\end{tabular}} & China     & 40.1       & 48.7       & 11.2    &/   & /        \\
                                                                          & U.S.      & 10.3       & 52.4       & 37.3     &/  & /        \\
\multirow{2}{*}{\begin{tabular}[c]{@{}c@{}}Qwen-Plus\end{tabular}} & China     & 34.7       & 53.2       & 10.6   &1.5    & 70.3\%       \\
                                                                          & U.S.      & 15.9      & 62.6       & 18.9   & 2.6   & 63.9\%       \\                                                                          
\multirow{2}{*}{\begin{tabular}[c]{@{}c@{}}GPT-4o mini\end{tabular}}   & China     & 45.5       & 39.3       & 10.3    & 0   & 79.0\%       \\
                                                                          & U.S.      & 17.7      & 47.3       & 35.0    &0   & 72.6\%       \\
                                                                          
\multirow{2}{*}{\begin{tabular}[c]{@{}c@{}}Qwen-Plus\\(PO)\end{tabular}} & China     & 50.1       & 38.9       & 11   &0    & 71.5\%       \\
                                                                          & U.S.      & 9.3       & 69.3       & 21.2   & 0.2   & 70.7\%       \\                                                                          
\multirow{2}{*}{\begin{tabular}[c]{@{}c@{}}GPT-4o mini\\(PO)\end{tabular}}   & China     & 45.6       & 43.5       & 10.9    & 0   & \textbf{88.4\%}       \\
                                                                          & U.S.      & 11.7       & 47.6       & 40.7    &0   & \textbf{85.9\%}       \\
\multirow{2}{*}{\begin{tabular}[c]{@{}c@{}}GPT-4o mini\\(FT)\end{tabular}}   & China     & 40.3       & 53.8       & 5.9    & 0   & 76.3\%       \\
                                                                          & U.S.      & 9.1       & 67.3       & 23.6    &0   & 75.0\%       \\

\multirow{2}{*}{\begin{tabular}[c]{@{}c@{}}GPT-4o mini\\(FT + PO)\end{tabular}}   & China     & 42.5       & 51.5       & 6.0    & 0   & 78.3\%       \\
                                                                          & U.S.      & 9.2       & 66.4      & 24.4    &0   & 73.9\%       \\
                                                                          
\hline
\end{tabular}
}
\label{sentiment-test} 
\end{table}

Based on these findings, we developed detailed operational guidelines for the LLM to ensure consistency and accuracy in classification. Ultimately, we comprehensively annotated 330,670 comments using the updated guidelines. Table \ref{sentiment} presents the results of the sentiment analysis of comments exchanged between Chinese and U.S. users regarding each other's countries. We categorized sentiments based on three labels: negative, Neutral, and positive.

\begin{table}[ht]
\centering
\caption{Sentiment distribution of China and U.S.}
\scalebox{0.8}{
\begin{tabular}{ccccccc}
\hline
      & Negative                                                   & Neutral                                                    & Positive                                                   & Total   \\ \hline
China & \begin{tabular}[c]{@{}c@{}}96,150\\ (45.55\%)\end{tabular} & \begin{tabular}[c]{@{}c@{}}91,322\\ (43.26\%)\end{tabular} & \begin{tabular}[c]{@{}c@{}}23,618\\ (11.19\%)\end{tabular} & 211,090 \\
U.S.  & \begin{tabular}[c]{@{}c@{}}14,439\\ (11.36\%)\end{tabular} & \begin{tabular}[c]{@{}c@{}}59,170\\ (46.55\%)\end{tabular} & \begin{tabular}[c]{@{}c@{}}53,508\\ (42.09\%)\end{tabular} & 127,117 \\ \hline
\end{tabular}
}
\label{sentiment} 
\end{table}

\subsection{Regression Analysis}

Since 2022, China’s \textit{Douyin} platform has mandated the disclosure of users' provincial location information alongside their comments, providing us with detailed data for analyzing regional emotional attitudes in China toward the United States. We linked the themes and sentiments of comments to their provincial IP addresses to study regional differences and explore the factors influencing these variations. To evaluate the likelihood of observing negative comments, we used a ordered logistic regression model~\cite{demaris1995tutorial}, where the dependent variable is the comment sentiment categorized into three levels: 1 representing negative, 2 representing neutral, and 3 representing positive. The provincial-level explanatory factors are sourced from China’s Seventh National Population Census. For further details, please refer to the results section. Our study focuses exclusively on data from China’s 31 provincial regions. Additionally, we applied a fixed effects design at the short video level to control for specific characteristics associated with each video. This approach allows for a more accurate analysis of the relationship between predictor variables and outcomes within video comments, minimizing the influence of content distribution algorithms on variations in exposure.

\section{RESULT}

\subsection{Topic Themes}
Figure \ref{topic-1} (1)-(2) present the word cloud distributions in both China and the U.S., based on the words' higher weighted log odds~\cite{monroe2008fightin}. The word cloud of the four dimensions, as shown in appendix \ref{appendix:A2}. Table \ref{topic-1} shows that Chinese video comments mainly focus on Power Transition Theory (30.32\%) and Complex Interdependence Theory (29.18\%), while U.S. comments emphasize Constructivism (48.79\%), reflecting differing concerns: China focuses on power dynamics and interdependence, while the U.S. highlights the role of ideas and perceptions in China-U.S. relations.

\begin{figure}[htp]
    \centering
    \includegraphics[width=8cm]{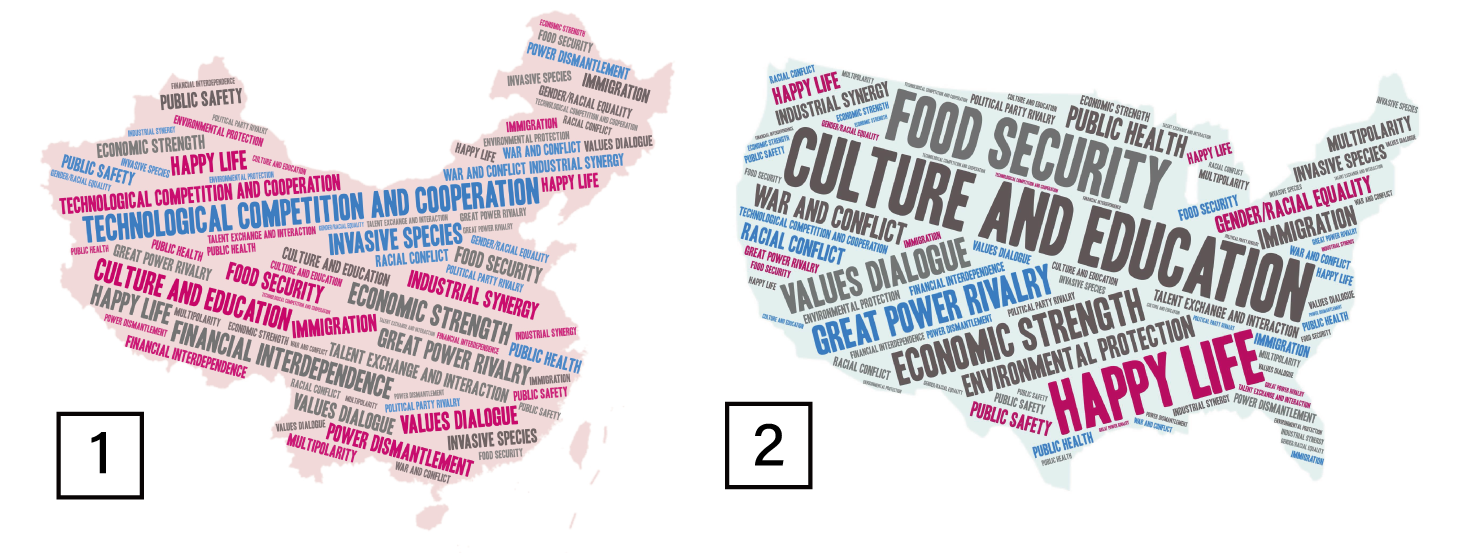}
    \caption{Word cloud map of (we translated the Chinese into English for display) (1) China; (2) U.S..}
    \label{topic-1}
\end{figure}

\begin{table}[ht]
\centering
\caption{Distribution of the four theoretical perspectives in the comments on China-U.S. relations}
\scalebox{0.7}{
\begin{tabular}{clcccc}
\hline
      & Total  & \begin{tabular}[c]{@{}c@{}}Power \\ Transition \\ Theory\end{tabular} & \begin{tabular}[c]{@{}c@{}}Complex \\ Interdependence\\ Theory\end{tabular} & Constructivism                                            & \begin{tabular}[c]{@{}c@{}}Global \\ Governance \\ Theory\end{tabular} \\ \hline
China & 211,090 & \begin{tabular}[c]{@{}c@{}}64,004\\ (30.32\%)\end{tabular}             & \begin{tabular}[c]{@{}c@{}}61,591\\ (29.18\%)\end{tabular}                   & \begin{tabular}[c]{@{}c@{}}49,744\\ (23.57\%)\end{tabular} & \begin{tabular}[c]{@{}c@{}}35,751\\ (16.93\%)\end{tabular}              \\
U.S.  & 127,117 & \begin{tabular}[c]{@{}c@{}}33,646\\ (26.47\%)\end{tabular}             & \begin{tabular}[c]{@{}c@{}}11,240\\ (8.84\%)\end{tabular}                    & \begin{tabular}[c]{@{}c@{}}62,023\\ (48.79\%)\end{tabular} & \begin{tabular}[c]{@{}c@{}}20,208\\ (15.90\%)\end{tabular}              \\ \hline
\end{tabular}
}
\label{topic-1}
\end{table}

\subsubsection{Power Transition Theory: Economic Strength and Values Dialogue}
The topic of ``Power Transition Theory'' is related to seven phrases: 1. Economic Strength, 2. Great Power Rivalry, 3. Power Dismantlement, 4. Values Dialogue, 5. War and Conflict, 6. Political Party Rivalry, 7. Racial Conflict. Among the U.S. and China, ``Economic Strength'' is the most frequently discussed, and comments typically reflect personal opinions. Chinese comments often express personal views through phrases like ``The Fed’s rate cut benefits China'' and ``The U.S. government’s debt problem is severe.'' U.S. comments include phrases like ``China’s rise is threatening the U.S. economy'' These comments highlight the global focus on the power transition between China and the U.S., centered on economic strength. 

\subsubsection{Complex Interdependence Theory: Mutual Dependence in Technology and Industry} The theme of ``Complex Interdependence Theory'' includes four phrases: 1. Technological Competition and Cooperation, 2. Financial Interdependence, 3. Industrial Synergy, and 4. Talent Exchange and Interaction. In China, ``Technological Competition and Cooperation'' is the most common, typically revolving around discussions such as ``If iPhone stops supporting updates for WeChat (a Chinese social media app), I won't use iPhone anymore'' and ``Huawei phones are far ahead.'' In the U.S., ``Industrial Synergy'' is more prevalent. Common phrases include ``Made in China is essential for U.S. manufacturing'' These comments highlight the mutual dependence between China and the U.S. in the fields of technology and economics. China emphasizes technological competition and cooperation across markets, while the U.S. focuses on industrial synergy, reflecting its reliance on Chinese products.

\subsubsection{Constructivism: Cultural Perceptions and Value Pursuits} From a Constructivist perspective, the key themes include: 1. Happy Life, 2. Culture and Education, 3. Immigration, 4. Public Safety, and 5. Gender/Racial Equality. Chinese discourse predominantly centers on the aspiration for a ``Happy Life,'' with emphasis on family well-being and future generations, as seen in comments like  ``I really envy your life.'' In contrast, U.S. discussions are dominated by ``Culture and Education,'' showcasing an appreciation and interest in Chinese cultural elements, such as ``I’m working hard to learn Chinese,'' and ``This Chinese song has a beautiful melody.'' These differing focuses highlight how cultural perceptions shape international relations. China prioritizes social welfare and prosperity, while the U.S. focuses on cultural exchange and education, reflecting constructivism's influence on bilateral relations and international dynamics.

\subsubsection{Global Governance Theory: Managing Global Challenges}
Includes five themes: 1. Invasive Species, 2. Food Security, 3. Public Health, 4. Environmental Protection, and 5. Multipolarity. In China, the primary focus is on managing invasive species, with discussions centered on the ecological impacts and control strategies, such as ``There are so many carp in the U.S., Chinese fishermen should help them'' and ``Using biological control to manage mosquitoes''. Conversely, U.S. discourse emphasizes ``Food Security,'' addressing sustainable food production and healthy eating choices with comments like ``This meat is lab-grown, very eco-friendly'' and ``I only eat organic food.'' These differing priorities reflect China and U.S. unique environmental and economic contexts within global governance.

\subsection{Sentiment in China and U.S.} 

\begin{figure}[ht]
    \centering
    \includegraphics[width= 8CM]{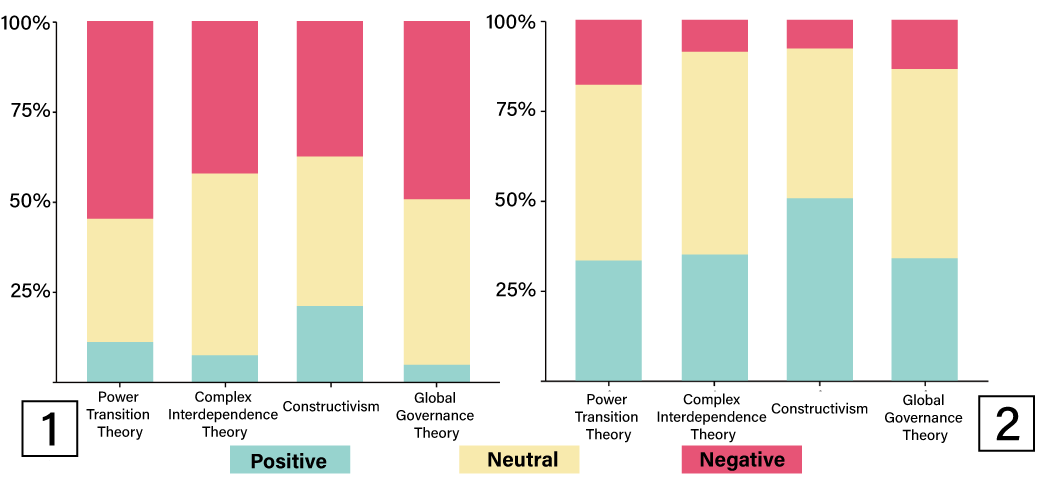}
    \caption{China-U.S. comments by sentiment (negative, positive, neutral) within four theoretical frameworks: (1) China’s comments on the U.S. (2) U.S. Comments on China.}
    \label{Sentiment by Theoretical}
\end{figure}

\begin{figure}[ht]
    \centering
    \includegraphics[width=6CM]{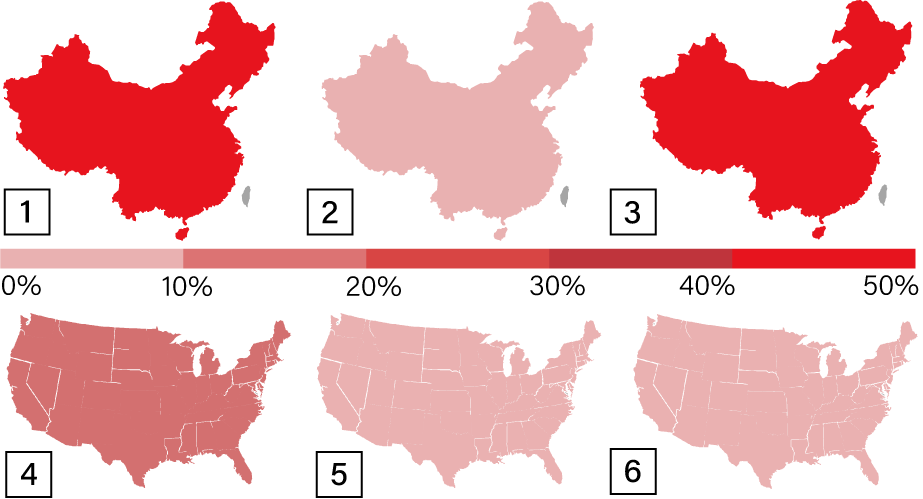}
    \caption{The levels of negative sentiment expressed by China and the U.S. towards each other are as follows: (1) China's overall negative sentiment, (2) China's hate sentiment, (3) China's criticism or complaints, (4) U.S. overall negative sentiment, (5) U.S. hate sentiment, and (6) U.S. criticism or complaints.}
    \label{Sentiment-2}
\end{figure}

We analyzed these figures according to theoretical dimension types. Figure \ref{Sentiment by Theoretical} illustrates the distribution of comments expressing negative, positive, and neutral sentiments across four theoretical dimensions in China (1) and the U.S. (2). As shown in Figures \ref{Sentiment-2} (1)-(3), in the Chinese sample, 45.55\% (96,150 comments) were categorized as ``negative sentiment towards the U.S.,'' including 3.82\% hate (8,068 comments) and 41.73\% criticism or complaints (88,082 comments). In contrast, in the U.S. sample, 11.36\% (14,439 comments) were categorized as ``negative sentiment towards China,'' including 2.63\% hate (3,340 comments) and 8.73\% criticism or complaints (11,099 comments), as shown in Figures \ref{Sentiment-2} (4)-(6). 

\begin{figure}[h]
    \centering
    \includegraphics[width=8CM]{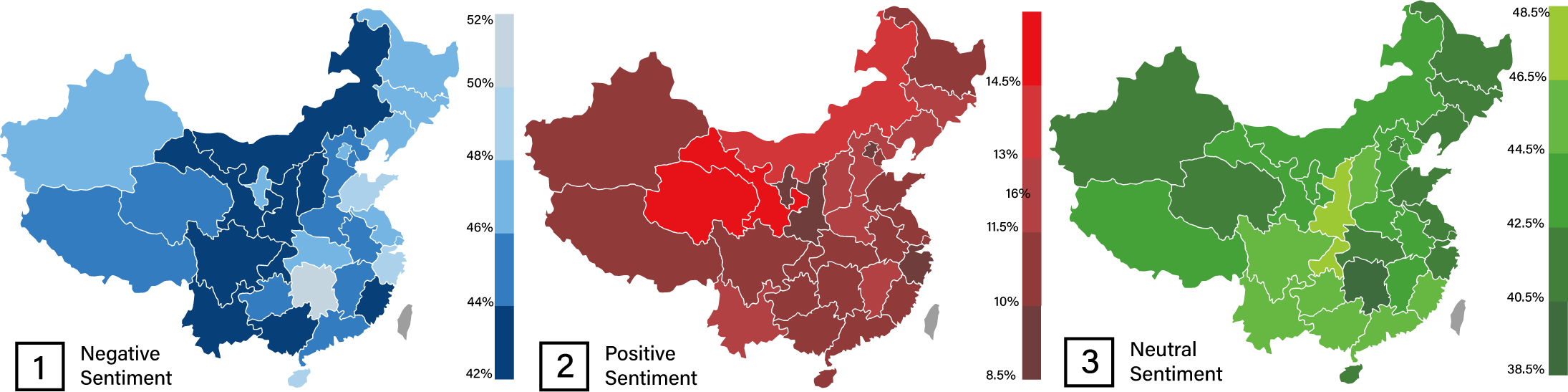}
    \caption{The levels of negative, positive, and neutral sentiment towards the U.S. across different provinces in China: (1) Negative. (2) Positive. (3) Neutral.}
    \label{Sentiment-1}
\end{figure}

Figure \ref{Sentiment-1} (1) (2) (3) display the levels of negative, positive, and neutral sentiment towards the U.S. across different provinces in China. As shown in Figure \ref{Sentiment-1} (2), the lowest levels of negative sentiment towards the U.S. were found in Shaanxi (42.13\%), followed by Chongqing (42.34\%) and Shanxi (42.51\%). In contrast, Hunan and Hainan exhibited the highest levels of negative sentiment towards the U.S., at 51.01\% and 48.38\%, respectively. 

\subsection{Levels of Negative Sentiment: A Regression Analysis in China}
We use regression analysis to examine variations in hate sentiment across four theoretical dimensions and identify thematic topics within short videos that attract higher levels of hate sentiment. Additionally, we explore how regional economic and cultural factors contribute to these sentiment differences, focusing on three key factors: (1) provincial GDP per capita (economic development), (2) the proportion of ethnic minorities, and (3) internet penetration rate. Control variables include the number of religious temples, the foreign population, and Starbucks stores, which reflect the influence of religious values \cite{halman1994religion}, multicultural interaction \cite{lebedeva2016intercultural}, and openness to global influences, respectively. Our regression results, shown in Table \ref{level-full}, indicate that Global Governance Theory is linked to significantly higher levels of negative sentiment compared to the other dimensions. This may be due to the complex and contentious nature of global governance issues, which often provoke emotional responses in social media discussions. Furthermore, Our results show that higher levels of economic development and internet penetration are associated with lower odds of comments being classified into more negative sentiment categories. In economically developed regions, globalization and exposure to diverse perspectives are more common. With increased international economic and cultural exchanges, users in these regions are more likely to adopt tolerant and open attitudes toward other countries and cultures. Furthermore, ethnic diversity is significantly associated with sentiment levels; regions with higher ethnic diversity exhibit lower odds of more negative sentiment. Frequent interactions with diverse cultural backgrounds and values may promote openness and inclusivity, encouraging users to respond with understanding and respect when encountering different cultural or value systems, rather than expressing hostility or negativity.

\begin{table}
\centering
\caption{Ordered Logistic Regression Model Predicting the Odds of ``Negative'' Comments}
\scalebox{0.8}{
\begin{tabular}{lc}
\hline
Variables                                                                                                             & Coefficient                                                                            \\ \hline
ComplexInterdependence vs GlobalGovernance                                                                             & -0.324***                                                                              \\
                                                                                                                      & (0.021)                                                                                \\
Constructivism vs GlobalGovernance                                                                                      & -0.254***                                                                              \\
                                                                                                                      & (0.021)                                                                                \\
PowerTransition vs GlobalGovernance                                                                                     & -0.246***                                                                              \\
                                                                                                                      & (0.021)                                                                                \\
GDP: middle vs bottom                                                                                                 & 0.001                                                                                  \\
                                                                                                                      & (0.014)                                                                                \\
GDPP: top vs bottom                                                                                                   & -0.042**                                                                                \\
                                                                                                                      & (0.020)                                                                                \\
Internet penetration rate                                                                                             & -0.282**                                                                               \\
                                                                                                                      & (0.138)                                                                                \\
Ethnic minority middle vs lowest                                                                                      & 0.009                                                                                  \\
                                                                                                                      & (0.014)                                                                                \\
Ethnic minority high vs lowest                                                                                        & -0.047***                                                                              \\
                                                                                                                      & (0.015)                                                                                \\
Religion                                                                                                              & -0.002                                                                                 \\
                                                                                                                      & (0.005)                                                                                \\
Foreign people                                                                                                        & 0.000                                                                                  \\
                                                                                                                      & (0.000)                                                                                \\
Starbucks                                                                                                             & -0.000                                                                                  \\
                                                                                                                      & (0.001)                                                                                \\
Video Fixed Effect                                                                                                                    & YES                                                                                \\
R2                                                                                                                    & 0.1551                                                                                \\
Number of Videos                                                                                                      & 1,142                                                                                   \\
Number of Comments                                                                                                    & 203,553                                                                                \\ \hline
\multicolumn{2}{c}{\begin{tabular}[c]{@{}c@{}}Data: \textit{Douyin}. \\ standard errors in parentheses.\\ * p ~\textless 0.1, ** p ~\textless 0.05, *** p ~\textless 0.01\end{tabular}}
\end{tabular}
}
\label{level-full}
\end{table}

\section{Discussion}

\textbf{China-U.S. Sentiment Divergence}
Our analysis reveals a significant emotional asymmetry in the responses from Chinese and U.S. commenters when discussing each other’s countries. Chinese comments largely express negative sentiments toward the U.S., focusing on perceived injustices, competitive threats, and issues of national pride. These views align with the realist perspective in international relations \cite{gries2004china}. In contrast, U.S. comments tend to focus more on cultural perceptions and value differences, rather than geopolitical competition. This reflects constructivist theory, which stresses that international relations are deeply shaped by ideas, identities, and social norms \cite{adler1997seizing}. This divergence highlights the distinct ways in which each country’s public sentiment is framed. For China, negative views toward the U.S. are driven by fears of national security and concerns about global power rivalry. U.S. perspectives, on the other hand, tend to focus more on cultural and ideological divides. This suggests that improving bilateral relations will require addressing each nation’s unique emotional and ideological concerns. For China, policies acknowledging national pride and addressing security issues could ease negative sentiments. In the U.S., fostering deeper cultural understanding and challenging stereotypes about China could bridge the perceptual gap.

\textbf{Regional Variations in China's Perceptions}
Our study reveals regional variations in emotional responses toward the U.S. within China. Provinces with lower GDP, fewer ethnic minorities, and limited internet access tend to express stronger negative emotions. This is due to limited media exposure, where official narratives often frame international relations as competitive, fueling nationalist sentiment \cite{gries2011patriotism}. In contrast, more developed regions with higher internet access and greater ethnic diversity encounter diverse perspectives, leading to more nuanced views of the U.S. Economic development and cultural diversity in these areas reduce sensitivity to global competition, mitigating negative sentiment. As China’s global role grows, addressing these regional differences is essential. Promoting media literacy, cross-cultural communication, and inclusive narratives could help reduce negative perceptions and foster more balanced public sentiment.

\textbf{Suggestions} \textit{Douyin} and \textit{TikTok} are important platforms for public discussion. The role these platforms play in content management means they have a significant influence in shaping mainstream narratives, which has sparked extensive discussions about content management and the formation of public opinion. Regarding \textit{Douyin} and \textit{TikTok}, the existing algorithms may to some extent impact the diversity of discourse and could promote the prominence of certain sentiment and narratives. This algorithm-based content distribution method may pose challenges for cross-cultural understanding, foster the formation of echo chambers, potentially exacerbate existing biases, and limit opportunities for people to encounter diverse viewpoints \cite{wei2024social,harris2023honestly}. As \textit{Douyin} and \textit{TikTok} shape public opinion and national identity, policymakers must understand the digital ecosystems guiding user cognition. Effective diplomacy relies on grasping regional socioeconomic contexts and algorithmic forces. Encouraging cross-cultural communication on these platforms can help reduce fragmentation and promote balanced dialogue.

\section{CONCLUSIONS}
China and the United States on social media platforms prompted us to investigate the discourses surrounding China-U.S. relations on \textit{Douyin} and \textit{TikTok}. Our study found that comments predominantly focused on themes such as ``Economic Strength,'' ``Technological Competition and Cooperation,'' ``Happy Life,'' and ``Invasive Species.'' Sentiment analysis showed that Chinese comments are more negative than U.S. ones. In China, higher GDP, greater ethnic diversity, and higher internet penetration are associated with more negative sentiments across regions. Overall, comments from areas with lower economic development, less ethnic diversity, and lower internet penetration tended to display more negative sentiments. We call for further research to delve into the underlying mechanisms and for strategies to enhance mutual understanding and reduce negative perceptions on social media.

\section{Limitations}

This study has several limitations, primarily stemming from the use of social media data from \textit{Douyin} and \textit{TikTok}. First, \textit{Douyin}’s geographic data is restricted to China’s provincial-level regions, which may obscure important local socio-economic factors and regional disparities relevant to discussions on China-U.S. relations. Additionally, \textit{Douyin}’s location-based recommendation algorithms could bias the dataset towards localized content, limiting the diversity of viewpoints. To enhance data representativeness, we employed IP proxies from multiple provinces. Secondly, the analysis of \textit{TikTok} is limited to accounts registered in the United States and English and Chinese language comments. Although some users may have relocated overseas, since the account registration remains in the U.S., we consider their comments to still reflect the attitudes of U.S. users toward China. We believe that the scale of our dataset helps to mitigate the potential interference caused by this minority scenario. 

Moreover, excluding non-English and non-chinese comments (such as those in Spanish) means the study primarily captures the perspectives of English-speaking and chinese-speaking \textit{TikTok} users in the U.S., potentially overlooking viewpoints from other linguistic communities within the country. The number of Chinese comments on tiktok is very small (less than 1\%), but we have retained the Chinese comments. Despite the large dataset size, it only includes users who actively participate in public comment sections, which may not accurately represent the broader user base \cite{inara2021let}. Previous research indicates that male users are more likely than female users to engage in online commentary, potentially leading to a gender bias in the perspectives analyzed \cite{kuchler2023gender}. While LLMs offer more nuanced labels that capture complex sentiments, their variable outputs can hinder replication \cite{gilardi2023chatgpt}. However, human annotators face similar challenges \cite{garcia2016challenges}, and in our study, the LLM maintained reasonably stable performance. Finally, the findings are specific to \textit{Douyin} and \textit{TikTok} users and should be cautiously generalized to the broader populations of China or the U.S. Online discussions on these platforms may not fully capture offline perceptions or broader social attitudes, as they are influenced by the distinct social and values environments of the two countries. 

We chose the number of temples as a measurable indicator of religious influence due to data availability, acknowledging its limitations in fully capturing the impact of religion and moral values. In the absence of comprehensive micro-level religious survey data in China, we used the number of religious institutions as a proxy, reflecting the extent of religious practice through their activities and memberships. Since most religious activities in China are organized by these institutions, their quantity provides a reasonable measure of religious influence in different regions \cite{wang2014does}. 

\textbf{ETHICAL STATEMENT} Our data is sourced from publicly available internet sources. We prioritize user privacy and security, especially when handling sensitive geo-location data. For comments from China, provincial geo-location information is used, but with each province housing over one million people, individual identification is unlikely. We do not infer user mobility or misuse the data; instead, we associate it with comments for analysis. The aim of our research is to understand public discussions on China-U.S. relations, not to stereotype or stigmatize any individual or country. We ensure that all data is anonymized, and only random samples are published. We ensured that all data was anonymized, and we only revealed a 10\% random sample with three variables: comments, topic labels, and sentiment labels. This project has been approved by our Institutional Review Board (IRB anonymous).

\bibliography{custom}

\clearpage
\appendix
\section{APPENDIX}
\subsection{40 Keywords for Search and Extraction}\label{appendix:A1}

\begin{table}[htbp]
\centering
\caption{\textbf{Keyword search and extraction between China and the United States}}
\scalebox{0.8}{
\begin{tabular}{lcc}
\hline
\multicolumn{1}{c}{\textbf{Type}} & \textbf{China (Douyin)} & \textbf{U.S. (TikTok)} \\ \hline
                                  & USA                     & China                  \\
                                  & American                & Chinese                \\
                                  & American Dream          & Chinese Dream          \\
                                  & American Society        & Chinese Society        \\
                                  & American Economy        & Chinese Economy        \\
                                  & American Development    & Chinese Development    \\
                                  & American Election       & Chinese Politics       \\
                                  & American Culture        & Chinese Culture        \\
                                  & American Travel         & China Travel           \\
                                  & US Interest Rate Cuts   & China Finance          \\
                                  & American Immigration    & Learn Chinese          \\
                                  & American Bankruptcy     & The Chinese Threat     \\
                                  & American Government     & Chinese Government     \\
                                  & Washington              & Beijing                \\
                                  & American Life           & China Life             \\
                                  & American Food           & Chinese Food           \\
                                  & Los Angeles             & Shanghai               \\
                                  & New York                & Guangzhou              \\
                                  & American Streets        & China Streets          \\
                                  & American Decline        & China Decline          \\ \hline
\end{tabular}
}
\label{A1}
\end{table}

\subsection{Word cloud map of four theoretical dimensions}\label{appendix:A2}

\begin{figure}
    \centering
    \includegraphics[width=8cm]{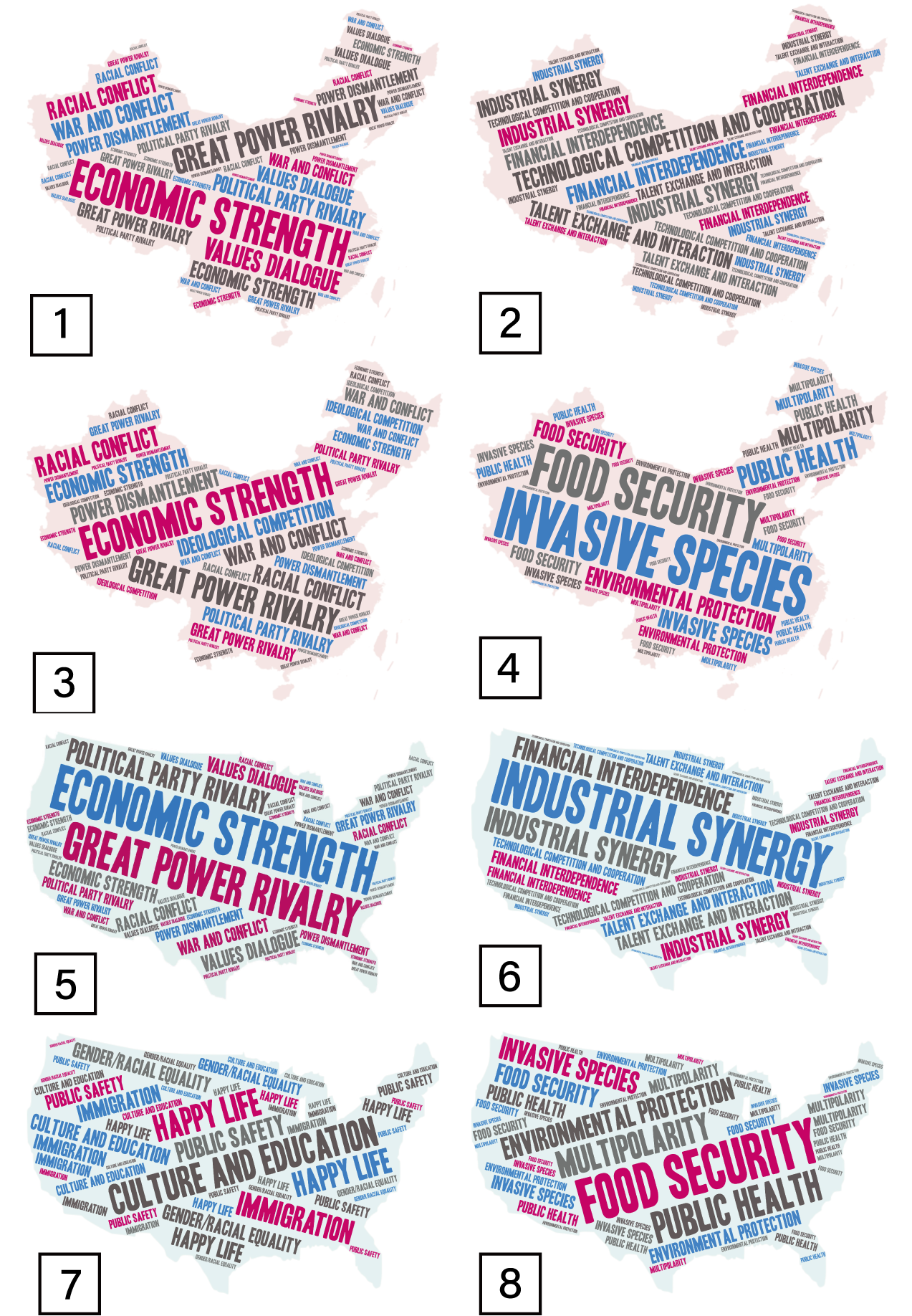}
    \caption{Word cloud map of (we translated the Chinese into English for display) (1) Power Transition Theory - 7 words in China; (2) Complex Interdependence Theory - 4 words in China; (3) Constructivism - 5 words in China; (4) Global Governance Theory - 5 words in China; (5) Power Transition Theory - 7 words in U.S.; (6) Complex Interdependence Theory - 4 words in U.S.; (7) Constructivism - 5 words in U.S.; (8) Global Governance Theory - 5 words in U.S..}
    \label{topic}
\end{figure}

\end{document}